\documentclass[lettersize,journal]{IEEEtran}

\usepackage[switch]{lineno}
\usepackage{array}
\usepackage{stfloats}
\usepackage{threeparttable}
\usepackage{verbatim}

\usepackage{cite}
\usepackage{amsmath,amssymb,amsfonts}
\usepackage{algorithmic}
\usepackage{graphicx}
\usepackage{subfigure}
\usepackage{textcomp}
\usepackage{xcolor}
\usepackage{colortbl}
\usepackage{float}
\usepackage{booktabs}
\definecolor{awesome}{rgb}{0.98, 0.66, 0.68}
\usepackage{multirow}
\usepackage{etoolbox}
\usepackage{setspace}
\usepackage{etoolbox}
\usepackage{algorithmic}
\usepackage{algorithm}

\def\BibTeX{{\rm B\kern-.05em{\sc i\kern-.025em b}\kern-.08em
    T\kern-.1667em\lower.7ex\hbox{E}\kern-.125emX}}
\usepackage{balance}
\begin{document}

\title{Task-Oriented Learning for Automatic EEG Denoising \\ 

\thanks{This work was supported in part by the National Key Research and Development Program of China under Grant 2023YFC2415100, in part by the National Natural Science Foundation of China under Grant 62373351, Grant 82327801, Grant 62073325, Grant 62303463, in part by the Chinese Academy of Sciences Project for Young Scientists in Basic Research under Grant No. YSBR-104, in part by the Beijing Natural Science Foundation under Grant F252068, Grant 4254107, in part by Beijing Nova Program under Grant 20250484813, in part by China Postdoctoral Science Foundation under Grant 2024M763535, in part by the Postdoctoral Fellowship Program of CPSF under Grant GZC20251170.}
\thanks{This work involved humans in its research. Approval of all ethical and experimental procedures and protocols was granted by the Ethics Committee of the Institute of Automation, Chinese Academy of Sciences.}
\thanks{Tian-Yu Xiang, Zheng Lei, Xiao-Hu Zhou, Xiao-Liang Xie, Shi-Qi Liu, Mei-Jiang Gui, Xin-Zheng Huang, and Xin-Yi Fu are with the State Key Laboratory of Multimodal Artificial Intelligence Systems, Institute of Automation, Chinese Academy of Sciences, Beijing 100190, China, and also with the School of Artificial Intelligence, University of Chinese Academy of Sciences, Beijing 100049, China (e-mail:xiangtianyu2021@ia.ac.cn; xiaohu.zhou@ia.ac.cn).}
\thanks{Hong-Yun Ou is with the School of life Science and Technology, Institute of Science Tokyo.}
\thanks{Zeng-Guang Hou is with State Key Laboratory of Multimodal Artificial Intelligence Systems, Institute of Automation, Chinese Academy of Sciences, Beijing 100190, China, also with the CAS Center for Excellence in Brain Science and Intelligence Technology, Beijing 100190, China, also with the School of Artificial Intelligence, University of Chinese Academy of Sciences, Beijing 100049, China, and also with the Joint Laboratory of Intelligence Science and Technology, Institute of Systems Engineering, Macau University of Science and Technology, Taipa, Macao, China (e-mail: zengguang.hou@ia.ac.cn).}
\thanks{$^\dagger$ Equal contribution: Tian-Yu Xiang, Zheng Lei, $*$ Corresponding authors: Xiao-Hu Zhou, Zeng-Guang Hou.}
}
\author{\IEEEauthorblockN{Tian-Yu Xiang$^\dagger$, Zheng Lei$^\dagger$, Xiao-Hu Zhou$^{*}$, \emph{Member, IEEE}, Xiao-Liang Xie, \emph{Member, IEEE}, Shi-Qi Liu}

\IEEEauthorblockN{Mei-Jiang Gui, Hong-Yun Ou, Xin-Zheng Huang, Xin-Yi Fu, Zeng-Guang Hou$^{*}$, \emph{Fellow, IEEE}}
}

\markboth{Journal of \LaTeX\ Class Files,~Vol.~xx, No.~x, xx~xxxx}%
{How to Use the IEEEtran \LaTeX \ Templates}\maketitle

\begin{abstract}

Electroencephalography (EEG) denoising methods typically depend on manual intervention or clean reference signals. This work introduces a task-oriented learning framework for automatic EEG denoising that uses only task labels without clean reference signals. EEG recordings are first decomposed into components based on blind source separation (BSS) techniques. Then, a learning-based selector assigns a retention probability to each component, and the denoised signal is reconstructed as a probability-weighted combination. A downstream proxy-task model evaluates the reconstructed signal, with its task loss supervising the selector in a collaborative optimization scheme that relies solely on task labels, eliminating the need for clean EEG references. Experiments on three datasets spanning two paradigms and multiple noise conditions show consistent gains in both task performance (accuracy: $2.56\%\uparrow$) and standard signal-quality metrics (signal-to-noise-ratio: $0.82$\,dB\,$\uparrow$). Further analyses demonstrate that the task-oriented learning framework is algorithm-agnostic, as it accommodates diverse decomposition techniques and network backbones for both the selector and the proxy model. These promising results indicate that the proposed task-oriented learning framework is a practical EEG denoising solution with potential implications for neuroscience research and EEG-based interaction systems.

\end{abstract}


\section{Introduction}

Decoding and analyzing brain activity from neuroimaging has long been pursued for its potential in rehabilitating individuals with disabilities and augmenting healthy function~\cite{penaloza2018bmi, willett2021high, 10663067}. Within available modalities, the electroencephalogram (EEG) is particularly attractive because it offers a favorable balance among cost, safety, and temporal resolution~\cite{sawangjai2019consumer}. However, EEG signals are intrinsically weak and highly susceptible to contamination. Noise in EEG recordings originates from three primary sources: physiological artifacts, instrumental factors, and environmental interference~\cite{mumtaz2021review, jiang2019removal}. These noises lead to a low signal-to-noise ratio (SNR) in EEG signals and obscure task-related information. 

These observations underscore the need for advanced denoising methods to improve EEG quality and increase EEG decoding reliability~\cite{mushtaq2024one}. Despite extensive efforts to capture high‑quality EEG data, the inherent combination of tightly coupled nonlinear noise and subtle neural components complicates the process of getting clean signals~\cite{mumtaz2021review}. Research spanning nearly five decades has yet to reach consensus on an optimal denoising algorithm for practical deployment~\cite{mumtaz2021review}. Moreover, a recent overview of EEG applications emphasized artifact detection and removal as one of the highest priorities in EEG signal processing~\cite{mushtaq2024one}.

Current EEG denoising techniques are classified into two categories: traditional signal‑processing–based methods and deep‑learning–based approaches. Traditional methods leverage signal‑processing techniques such as filtering~\cite{AdaptiveFilters2007}, regression~\cite{Regression-Base2009validation}, and BSS (e.g., independent component analysis (ICA)~\cite{MNE-ICALabel2022}, principal component analysis (PCA)~\cite{de2007denoising}, canonical correlation analysis (CCA)~\cite{CCA2009}). Filtering and regression approaches typically require handcrafted features derived from statistical analysis, whereas BSS approaches usually depend on manual selection of components to exclude noise. This reliance on handcrafted features and manual component selection limits applicability in practical scenarios.

Deep-learning–based approaches implement an end-to-end pipeline that requires neither pre-defined indices nor manual intervention. These methods can be categorized into two main types: autoencoder-based~\cite{zhang2021eegdenoisenet, NovelCNN2021novel, Denoiseformer2023denosieformer, EEGDNet2022, DuoCL2022} and generative adversarial network (GAN)-based models~\cite{Denoise-GAN2022denoising, EEGANet2021, yin2023gan}. Autoencoder-based models learn a direct mapping from noisy to clean EEG signals via architectures designed for EEG representation extraction, such as convolutional neural networks (CNN)~\cite{zhang2021eegdenoisenet, NovelCNN2021novel}, transformers~\cite{Denoiseformer2023denosieformer, EEGDNet2022}, and recurrent neural networks (RNN)~\cite{DuoCL2022}. GAN-based methods comprise a generator \(G\) and a discriminator \(D\)~\cite{Denoise-GAN2022denoising, EEGANet2021, yin2023gan}; \(G\) produces candidate clean EEG signals, while \(D\) evaluates whether an input signal is real or generated. Training proceeds adversarially to optimize \(G\) such that \(D\) cannot distinguish generated denoised signals from true clean EEG data.

However, autoencoder-based methods require paired noisy and clean EEG recordings~\cite{zhang2021eegdenoisenet, NovelCNN2021novel, Denoiseformer2023denosieformer, EEGDNet2022, DuoCL2022}, and GAN-based methods depend on a representative set of clean EEG data to characterize its distribution~\cite{Denoise-GAN2022denoising, EEGANet2021, yin2023gan}. Acquiring high-quality clean EEG data is inherently challenging. Moreover, because these methods are trained to mimic clean signals, they may introduce artifacts or omit task-relevant information. Although deep-learning–based approaches facilitate automated denoising, opportunities remain to optimize these studies.

To address the limitations noted above, a task-oriented learning framework for automatic EEG denoising is proposed. The approach combines classical BSS signal processing with deep learning. Raw EEG is first decomposed into components based on BSS algorithms. Then, a learning-based selector identifies informative components and reconstructs a denoised signal as a probability-weighted combination. Selector training is guided by a proxy-task model in a collaborative optimization scheme, enabling selection of task-relevant components without requiring component-level labels. In contrast to deep learning-based methods that directly generate clean signals from noisy inputs, the proposed task-oriented learning framework takes advantage of decomposition components derived from the BSS algorithms, avoiding generating artifacts. By integrating signal-processing and deep-learning techniques, the denoising process operates without clean reference signals, preserves task-related information, and performs EEG denoising automatically.

The primary contributions of this study are as follows:

\begin{itemize}
\item A task-oriented learning framework for automatic EEG denoising is introduced that trains using only task-related labels, without clean reference signals.
\item A collaborative optimization scheme is developed in which a proxy-task model guides the selector to emphasize informative components within EEG signals.
\item Empirical evaluation on three datasets and two paradigms demonstrates improvements in signal quality (signal-to-noise-ratio: $0.82$\,dB\,$\uparrow$) and task-relevant metrics (accuracy: $2.56\%\uparrow$); compatibility with diverse BSS methods and learning-based backbones confirms this task-oriented learning framework is algorithm-agnostic.
\end{itemize}

The remainder of this paper is organized as follows. Section II reviews related work. Section III details the proposed approach. Section IV presents the experimental results and ablation studies. Section V discusses the effectiveness of the task-oriented learning framework from the perspective of signal properties. Section VI concludes the paper.

\section{Related Work}

\subsection{Traditional Signal-processing–based Methods}

EEG denoising is a fundamental signal-processing problem. Numerous approaches based on classical signal processing have been developed, among which BSS is a representative technique~\cite{MNE-ICALabel2022, de2007denoising, gong2017improved, CCA2009, EMD-ICA2015removal, EMD-CCA2017, ASR2019evaluation}. It decomposes EEG signals into constituent components via matrix decomposition methods such as ICA~\cite{MNE-ICALabel2022}, PCA~\cite{de2007denoising}, and CCA~\cite{CCA2009}. Although decomposition itself is straightforward, identifying which components carry task-relevant information while excluding noise remains challenging. Although automated component selection frameworks have been proposed~\cite{MNE-ICALabel2022, ASR2019evaluation}, many implementations and analyses still depend on manual selection.

Another traditional signal-processing approach treats EEG signals as composite and applies techniques such as wavelet transform~\cite{WT2016method} and adaptive filtering~\cite{AdaptiveFilters2007} for denoising~\cite{WT2016method, AdaptiveFilters2007, Regression-Base2009validation}. These methods rely on predefined, handcrafted rules or characteristics. The traditional signal-processing-based methods yield interpretable operations that are applied directly to the original recordings and, when properly configured, do not introduce spurious artifacts. Nevertheless, an automated end-to-end pipeline for EEG denoising remains in high demand.

\subsection{Deep-learning-based Methods}

Deep-learning–based approaches offer an automated framework for EEG denoising. These approaches are categorized into autoencoder-based and GAN-based methods. Autoencoder-based models are trained to map noisy EEG inputs to denoised outputs, typically via architectures capable of capturing spatial–temporal representations such as CNN~\cite{1D-ResCNN2020, NovelCNN2021novel, zhang2021eegdenoisenet, DeepSeparator2022embedding, SD-Net2023segmentation}, RNN~\cite{DuoCL2022, EEGIFNet2023dual}, and transformer~\cite{EEGDNet2022, Denoiseformer2023denosieformer, EEGDiR2024eegdir, STFNet2024, ART2025augmenting}. Training is generally performed by minimizing the mean squared error between the denoised output and a reference clean EEG signal. 

GAN-based methods do not perform direct regression from noisy to clean signals~\cite{yin2023gan, EEGANet2021, Denoise-GAN2022denoising}. Instead, a generator produces candidate clean signals while a discriminator evaluates whether a given signal is real or generated. The adversarial training objective encourages the generator to approximate the distribution of clean EEG signals rather than memorize individual examples, yielding outputs that are similar to the true clean signals. 

Despite the automation provided by both autoencoder- and GAN-based methods, their reliance on clean EEG data during training poses a significant challenge due to the difficulty of obtaining high-quality references. Moreover, because the denoising process is typically implemented with deep neural networks, interpretability is limited, and there is a risk of introducing new artifacts or discarding task-relevant information.

\begin{figure*}[tb]
    \centering
      \includegraphics[width=7in]{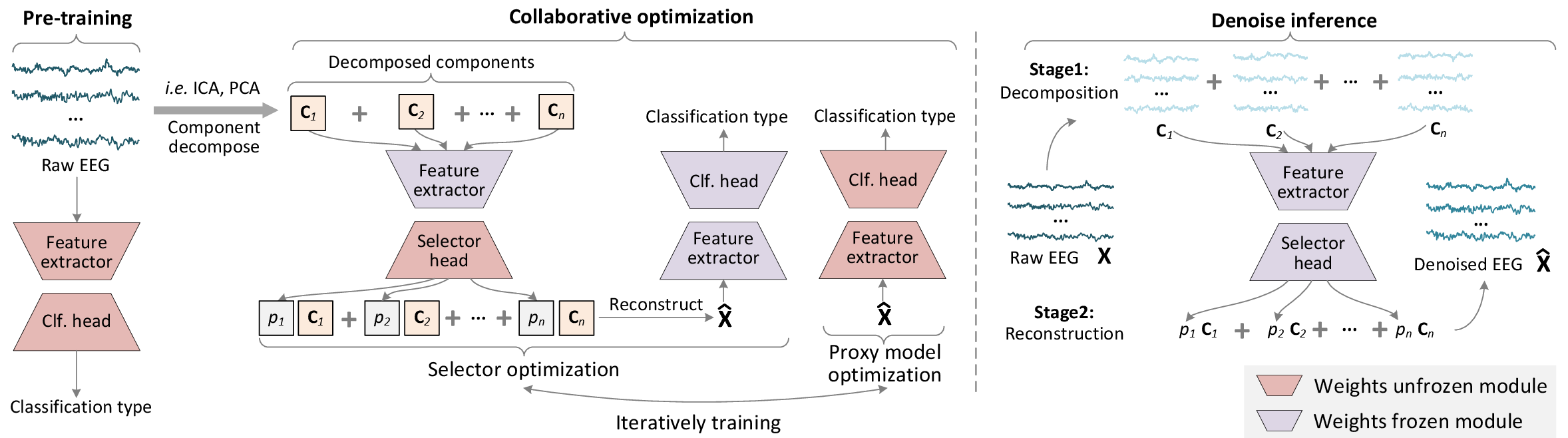}
        \caption{Overview of the proposed task-oriented learning framework for automatic EEG denoising. (i) Pre-training: the feature extractor and classification (Clf.) head are trained on a supervised proxy classification task using raw EEG. (ii) Collaborative Optimization: EEG is decomposed into components (e.g., ICA/PCA); the selector predicts component-retention probabilities and is optimized collaboratively with the proxy-task model. (iii) Denoise inference: the trained selector alone weights task-relevant components to reconstruct the denoised EEG.} 
     \label{fig:pipeline}
 \end{figure*}

\section{Method}

\subsection{Task-Oriented Learning Framework}
The proposed task-oriented learning framework for automatic EEG denoising is model-agnostic (shown in Fig.~\ref{fig:pipeline}), applied with different BSS decomposition algorithms and network structures. Raw EEG data $X$ is first decomposed into $N$ components $\{C_i\}_{i=1}^{N}$ using a BSS algorithm:

\begin{equation}
    \mathbf{X} = g(\mathbf{X}) = \sum_{i=1}^{N} \mathbf{C}_i
    \label{Eq.decomp}
\end{equation}
where $g(\cdot)$ denotes the decomposition operator. Each component $\mathbf{C}_i$ is then assigned a selection score $p_i$ by a selector function $f_{\mathrm{s}}(\cdot)$:
\begin{equation}
    p_i = f_{\mathrm{s}}\bigl(\mathbf{C}_i\bigr),
    \label{eq:recon0}
\end{equation}
where $p_i\in[0,1]$ represents the probability that $\mathbf{C}_i$ is informative rather than noisy. The denoised EEG signal $\hat{\mathbf{X}}$ is reconstructed as a weighted sum of components:
\begin{equation}
    \hat{\mathbf{X}} = \sum_{i=1}^{N} p_i\,\mathbf{C}_i.
    \label{eq:recon1}
\end{equation}
as the label for components is unknown, a proxy downstream task is defined to guide the training process of the selector. Let $\mathbf{y}$ denote the ground-truth label and $f_{\mathrm{p}}(\cdot)$ the proxy-task model. The selector loss is given by
\begin{equation}
    \mathcal{L} = c\bigl(f_{\mathrm{p}}(\hat{\mathbf{X}}),\,\mathbf{y}\bigr),
    \label{Eq:loss}
\end{equation}
where $c(\cdot,\cdot)$ is a chosen loss function. Minimizing $\mathcal{L}$ encourages retention of components that enhance proxy-task performance and suppression of task-irrelevant noisy components. 

\subsection{EEG Decomposition}

Although various decomposition methods exist for EEG analysis, the result can generally be expressed as
\begin{equation}
    \mathbf{X} = g(\mathbf{X}) = \mathbf{A}\,\mathbf{B}^\top + \boldsymbol{\epsilon},
    \label{eq:decomp0}
\end{equation}
where $\mathbf{X}\in\mathbb{R}^{C\times T}$ denotes the raw EEG signal with $C$ electrodes and $T$ time points, $\mathbf{A}\in\mathbb{R}^{C\times M}$ and $\mathbf{B}\in\mathbb{R}^{T\times M}$ are factor matrices with $M\le C$, and $\boldsymbol{\epsilon}$ is the residual matrix.  
Let $\mathbf{a}_i\in\mathbb{R}^{C\times1}$ and $\mathbf{b}_i\in\mathbb{R}^{T\times1}$ denote the $i$th columns of $\mathbf{A}$ and $\mathbf{B}$. Then Eq. \ref{eq:decomp0} admits the component-wise form:
\begin{equation}
\begin{aligned}
    \mathbf{X}
    &= \sum_{i=1}^{M}\mathbf{a}_i\,\mathbf{b}_i^\top + \boldsymbol{\epsilon}
    = \sum_{i=1}^{M} \mathbf{C}_i + \mathbf{C}_{M+1}
    = \sum_{i=1}^{N} \mathbf{C}_i,
\end{aligned}
\label{eq:decomp1}
\end{equation}
where
\[
    \mathbf{C}_i = \mathbf{a}_i\,\mathbf{b}_i^\top \quad (i=1,\dots,M), 
    \quad
    \mathbf{C}_{M+1} = \boldsymbol{\epsilon}, 
    \quad
    N = M + 1.
\]
each component $\mathbf{C}_i$ is the outer product of a spatial pattern $\mathbf{a}_i$ and a temporal pattern $\mathbf{b}_i$, so that each row of $\mathbf{C}_i$ encodes the signal distribution across electrodes for that component.

Hence, the operator $g(\cdot)$ in Eq.~\ref{Eq.decomp} can be treated as two steps: the initial matrix factorization in Eq.~\ref{eq:decomp0} and the post‐processing into accumulated spatial-temporal components in Eq.~\ref{eq:decomp1}.

\subsection{Collaborative Optimization}

\subsubsection{Optimization Process}

\begin{algorithm}[!tb]
	\caption{\textbf{: }Training Process}
	\begin{algorithmic}[1]
    \REQUIRE Dataset $D = \{(\mathbf{X}_{j}, \mathbf{y}_{j})\}$, epochs $E$, batch size $S$
    \ENSURE Selector $f_{s}$ predicts the retention probability of each component
    \STATE \textbf{— Pre-training stage—}
    \STATE Initialize the proxy model ($f_{p}$) (Eq.~\ref{Eq:proxy model})
    \STATE Pre‑train $f_{p}$ on $D$ using supervised learning
    \STATE Initialize selector head $h_{s}$ and shared $q$ to form $f_s$ (Eq.~\ref{Eq:selector})
    \STATE \textbf{— Collaborative optimization stage —}
    \FOR{epoch = $1$ to $E$}
        \REPEAT
            \STATE Sample a batch $B = \{(\mathbf{X}_{k},\mathbf{y}_k)\}_{k=1}^{S}$ from $D$.
            \STATE Decompose each $\mathbf{X}\in B$ via $g(\cdot)$ (Eqs.~\ref{eq:decomp0},\ref{eq:decomp1})
            \STATE Reconstruct EEG signals (Eqs.~\ref{eq:recon0},\ref{eq:recon1})
            \STATE \textbf{— Update Selector —}
            \STATE Unfreeze $h_{s}$; freeze $q$, $h_{p}$
            \STATE Optimize $h_{s}$ with AdamW based on Eq.~\ref{Eq:loss}
            \STATE \textbf{— Update Proxy Model —}
            \STATE Unfreeze $q$, $h_{p}$; freeze $h_{s}$
            \STATE Optimize $f_{p}$ with AdamW based on Eq.~\ref{Eq:loss}
        \UNTIL{All samples in $D$ have been sampled.}
    \ENDFOR
\end{algorithmic}
	\label{alg:opt}
\end{algorithm}

The optimization process involves two models: a selector, responsible for identifying task-relevant components, and a model for the proxy‑task. Since component labels are unavailable, guidance for the selector is provided by performance on the proxy-task. This design is motivated by the assumption that components related to the task contribute useful information, whereas those task-irrelevant ones introduce noise. Therefore, the performance of reconstructed EEG signals on the proxy-task is adopted as the evaluation metric for the selector model.

The selector ($f_s$) comprises two modules:
\begin{equation}
    f_s = h_s \circ q
    \label{Eq:selector}
\end{equation}
where $q$ denotes the shared feature extractor with the proxy-task model and $h_s$ represents the selector head, which outputs the probability that an input component is not noisy. Although the selector and proxy downstream classification tasks differ, both depend on the same extracted task‑related representations. Thus, the feature extractor is shared by these two models. The proxy‑task model ($f_p$) is defined as:
\begin{equation}
    f_p = h_p \circ q
    \label{Eq:proxy model}
\end{equation}
where $h_p$ denotes the classification head for the downstream task. Both $h_s$ and $h_p$ are implemented as fully connected layers that map the extracted features from the EEG or the decomposed components to output probabilities.

The detailed optimization procedure is presented in Algorithm~\ref{alg:opt}, which consists of two stages: pre‑training and collaborative optimization. During pre‑training, the feature extractor $q$ is initialized via supervised learning to establish a robust shared representation for both the selector and proxy-task model. This process lets the selector focus on the needs of learning to discriminate features related to the task.

The second stage implements an iterative collaborative optimization between the selector and the proxy-task model. In each iteration, the selector is optimized under the supervision of proxy‑task performance, thereby selecting components that enhance downstream task accuracy. Subsequently, the proxy-task model is refined by minimizing the classification loss computed on the reconstructed EEG signals and their corresponding labels. This collaborative optimization scheme guides the selector toward retaining task‑relevant components and discarding noise, thereby achieving EEG denoising.

\subsubsection{Convergence Analysis}

\label{Sec. Convergence Analysis}

To analyze the convergence of the collaborative optimization, the following hypotheses are adopted:
\begin{itemize}
    \item \noindent\textbf{H1}: $\inf \mathcal{L} > -\infty.$
    \item \noindent\textbf{H2}: The loss $c(\cdot,\cdot)$ and $h_p$, $q$, and $h_s$ are continuously differentiable with Lipschitz–continuous gradients on a bounded domain; $\hat{\mathbf{X}}$ is continuously differentiable with respect to the model parameters.
    \item \noindent\textbf{H3}: Gradients are bounded and the variance of the mini-batch stochastic noise is bounded; features lie in a bounded set.
    \item \noindent\textbf{H4}: Each module update does not increase the loss, either in full-batch form or in expectation in the stochastic setting. 
\end{itemize}

In the implementation, these hypotheses are encouraged as follows. \textbf{H1} holds because $\mathcal{L}$ is a cross-entropy loss and is therefore bounded below by zero. To promote \textbf{H2}, ELU activations are used, max-norm constraints are imposed on layer weights, and gradient clipping is applied. For \textbf{H3}, gradient clipping yields bounded gradients, while the feature extractor operating with bounded weights on finite-energy inputs ensures bounded feature vectors. Regarding \textbf{H4}, AdamW with the AMSGrad variant and a cosine learning-rate decay is employed to stabilize updates so that the loss is non-increasing in expectation.

Based on the above hypotheses, a discussion regarding the convergence is carried out. Define the collaborative objective:

\begin{equation}
\label{eq:joint_obj}
\mathcal{L}\left(\theta_s,\theta_p\right)\;\triangleq\; c\left(f_p\left(\hat{\mathbf{X}}\left(\theta_s,\theta_p\right)\right),\,\mathbf{y}\right),
\end{equation}
where $\theta_s$ is the parameters of $h_s$, and $\theta_p$ collects the parameters of $q$ and $h_p$. Under the hypotheses, $\mathcal{L}$ is block-wise continuously differentiable, and for fixed $\theta_p$ (or $\theta_s$), $\nabla_{\theta_s}\mathcal{L}$ (or $\nabla_{\theta_p}\mathcal{L}$) is Lipschitz continuous on a bounded domain.

The update can be considered iteratively, step A, $\exists \ \ \alpha_s$:
\begin{equation}
    \theta_s^{t+1}=\theta_s^{t}-\alpha_s\,\nabla_{\theta_s}\mathcal{L}(\theta_s^{t},\theta_p^{t}),\quad 
\alpha_s\in(0,\,2/\beta_s)
\end{equation}
and step B, $\exists \ \ \alpha_p$:
\begin{equation}
    \theta_p^{t+1}=\theta_p^{t}-\alpha_p\,\nabla_{\theta_p}\mathcal{L}(\theta_s^{t+1},\theta_p^{t}),\quad 
\alpha_p\in(0,\,2/\beta_p)
\end{equation}
where $\alpha_s$ and $\alpha_p$ is the learning rate for each step, $\beta_s,\beta_p>0$, $\mathcal{L}$ is $\beta_s$-smooth in $\theta_s$ with $\theta_p$ fixed and $\beta_p$-smooth in $\theta_p$ with $\theta_s$ fixed (block-wise Lipschitz gradients), and $t$ is the number of the current iteration. In step A, $\theta_p$ is treated as the constant, while in step B, $\theta_s$ is constant.

By the descent lemma for a $\beta_s$-smooth function and Step~A,
\begin{equation}
\label{eq:descentA}
\begin{aligned}
\mathcal{L}(\theta_s^{t+1},\theta_p^{t})
&\le \mathcal{L}(\theta_s^{t},\theta_p^{t})
 + \big\langle \nabla_{\theta_s}\mathcal{L}(\theta_s^{t},\theta_p^{t}),\,\theta_s^{t+1}-\theta_s^{t}\big\rangle \\
& + \tfrac{\beta_s}{2}\|\theta_s^{t+1}-\theta_s^{t}\|^2\\
&=\mathcal{L}(\theta_s^{t},\theta_p^{t}) - \Big(\alpha_s-\tfrac{\beta_s}{2}\alpha_s^2\Big)\,
\big\|\nabla_{\theta_s}\mathcal{L}(\theta_s^{t},\theta_p^{t})\big\|^2.
\end{aligned}
\end{equation}
Because $\alpha_s\in(0,2/\beta_s)$ or satisfies an Armijo condition, the coefficient $\mu_s\triangleq \alpha_s-\tfrac{\beta_s}{2}\alpha_s^2$ is strictly positive, yielding
\begin{equation}
\label{eq:A-final}
\mathcal{L}(\theta_s^{t+1},\theta_p^{t})
\le \mathcal{L}(\theta_s^{t},\theta_p^{t}) - \mu_s
\big\|\nabla_{\theta_s}\mathcal{L}(\theta_s^{t},\theta_p^{t})\big\|^2.
\end{equation}
Similarly, holding $\theta_s^{t+1}$ fixed in Step~B and using $\beta_p$-smoothness in $\theta_p$,
\begin{equation}
\label{eq:descentB}
\mathcal{L}(\theta_s^{t+1},\theta_p^{t+1})
\le \mathcal{L}(\theta_s^{t+1},\theta_p^{t}) - \mu_p
\big\|\nabla_{\theta_p}\mathcal{L}(\theta_s^{t+1},\theta_p^{t})\big\|^2
\end{equation}
\begin{equation}
    \mu_p\triangleq \alpha_p-\tfrac{\beta_p}{2}\alpha_p^2>0.
\end{equation}
Combining Eq. \ref{eq:A-final} and Eq. \ref{eq:descentB} gives

\begin{equation}
\begin{aligned}
\mathcal{L}(\theta_s^{t+1},\theta_p^{t+1})
&\le \mathcal{L}(\theta_s^{t},\theta_p^{t})
 - \mu_s\big\|\nabla_{\theta_s}\mathcal{L}(\theta_s^{t},\theta_p^{t})\big\|^2 \\
&  - \mu_p\big\|\nabla_{\theta_p}\mathcal{L}(\theta_s^{t+1},\theta_p^{t})\big\|^2.
\end{aligned}
\end{equation}

Thus $\mathcal{L}(\theta_s^{t+1},\theta_p^{t+1}) \le \mathcal{L}(\theta_s^{t},\theta_p^{t})$ holds for all $t$. Moreover, by \textbf{H1}, $\mathcal{L}$ has a finite lower bound, hence the monotone bounded sequence $\{\mathcal{L}(\theta_s^{t},\theta_p^{t})\}$ converges.

\subsection{EEG Denoising Inference}

Following training, raw EEG signals are denoised in two stages (Fig.~\ref{fig:pipeline}, right). Stage 1 (Decomposition): operator \(g(\cdot)\) decomposes each raw EEG signal into a group of components. Stage 2 (Reconstruction): the selector predicts the retention probability of each component, and the denoised EEG signal is reconstructed by combining components weighted by the corresponding predicted probabilities for downstream applications.

\section{Experiments}

The experiments are designed to answer the following questions:
\begin{itemize}
    \item Does the framework work for different paradigms?
    \item Does the framework work for different noises?
    \item Does the framework achieve a competitive performance with the previous EEG denoising algorithms?
    \item Does the framework retain task-related information during the denoising process?
\end{itemize}

\subsection{Dataset Description}

Three EEG datasets from two paradigms are used in this study, including two publicly available datasets~\cite{Cho2017, nakanishi2015comparison} and one self-collected dataset~\cite{xiang2023learning}.
\begin{itemize}
    \item SSVEP dataset: Nakanishi et al.~\cite{nakanishi2015comparison} provides a twelve-class SSVEP dataset containing 9 subjects, with 15 EEG trials of 4.15\,s per class for each subject. Data are acquired with an 8-channel EEG system at 256\,Hz.
    \item Motor imagery dataset: Cho et al.~\cite{Cho2017} presents a two-class motor imagery dataset comprising 52 subjects, with 100 EEG trials of 3\,s per class for each subject. Data are recorded using a 64-channel EEG acquisition system at 512\,Hz and downsampled to 160\,Hz for the experiments.
    \item Motor execution dataset: Xiang et al.~\cite{xiang2023learning, 10722863} is a self-collected four-class motor execution dataset including 11 subjects, with 100 EEG trials of 4\,s per class for each subject. Data are recorded using a 32-channel EEG system at 128\,Hz.
\end{itemize}

Because motor imagery and motor execution share the same underlying physiological mechanism: event-related desynchronization/synchronization (ERD/ERS) in the sensorimotor cortex~\cite{pfurtscheller1999event}. These two paradigms are grouped, whereas SSVEP is treated separately.

\subsection{Experimental Details}

\subsubsection{Noise Considered in this Study}

\paragraph{Simulated Physiological Artifacts}

Following prior work on simulated physiological artifacts~\cite{zhang2021eegdenoisenet}, electrooculogram (EOG) and electromyogram (EMG) signals are linearly combined with the EEG signals to emulate irregular eye blinks and muscle activity. To synthesize varying noise levels, signal-to-noise ratio (SNR) values are drawn uniformly from \([-5,5]\)\,dB, consistent with previous EEG noise studies~\cite{ROMERO2008348}. Because eye blinks and muscle artifacts typically do not affect the signal measured in the occipital cortex area associated with the SSVEP paradigm~\cite{nakanishi2015comparison}, these simulated physiological artifacts are excluded from the SSVEP dataset.

\paragraph{Inherent Instrumental and Environmental Noise}

Raw EEG signals acquired during measurement are inherently contaminated by instrumental and environmental noise, which is inseparable from the neural activities.

\subsubsection{Evaluation Metrics}

The denoising algorithms are evaluated from two perspectives:
\begin{itemize}
    \item The extent to which the denoising process preserves task-related information in the EEG signal.
    \item The effectiveness of the denoising process in reducing noise, i.e., improving signal quality.
\end{itemize}

\paragraph{Task-related Information}
Task-related information is assessed via proxy-task performance. Since the experimental paradigms, including SSVEP and motor imagery/execution, are classification problems. Standard classification metrics (accuracy, precision, recall, and F1 score) are reported to quantify the extent to which task-relevant information is preserved in the EEG signal after denoising.

\paragraph{Signal Quality Evaluation}

Quality evaluation compares denoised signals to reference signals. Since clean EEG signals are unavailable, following prior work~\cite{zhang2021eegdenoisenet}, raw EEG recordings are employed as the reference. Denoising performance is assessed by comparing artifact-contaminated signals after denoising with the original raw EEG. Thus, this evaluation addresses only the removal of simulated physiological artifacts, not the inherent instrumental or environmental noise.

Two evaluation metrics are employed. The mean squared error (MSE) quantifies the discrepancy between the denoised signal and the reference signal:
\begin{equation}
    \mathrm{MSE}(\hat{\mathbf{X}}, \mathbf{X}_{\mathrm{ref}}) = \frac{1}{CM} \sum_{i=1}^{C} \sum_{j=1}^{M} \left(\hat{\mathbf{X}}(i,j) - \mathbf{X}_{\mathrm{ref}}(i,j)\right)^2,
\end{equation}
where $\mathbf{X}_{\mathrm{ref}}$ denotes the reference signal, $\hat{\mathbf{X}}(i,j)$/$\mathbf{X}_{\mathrm{ref}}(i,j)$ is the $i$th electrode and the $j$th time point for the denoised signal/reference signal. The signal-to-noise ratio is defined in decibels (dB) as the ratio of the reference signal magnitude to the residual (error) magnitude:
\begin{equation}
    \mathrm{SNR}(\hat{\mathbf{X}}, \mathbf{X}_{\mathrm{ref}}) = 20 \log \left( \frac{\mathrm{RMS}(\mathbf{X}_{\mathrm{ref}})}{\mathrm{RMS}(\hat{\mathbf{X}} - \mathbf{X}_{\mathrm{ref}})} \right),
\end{equation}
with
\begin{equation}
    \mathrm{RMS}(\mathbf{X}) = \sqrt{ \frac{1}{CM} \sum_{i=1}^{C} \sum_{j=1}^{M} \mathbf{X}(i,j)^2 }.
\end{equation}
a lower MSE and a higher SNR compared with the reference signals indicate more effective denoising.

\subsubsection{Experimental Settings}

Experiments follow a per-subject evaluation protocol with 5-fold cross-validation. For each subject, a model is trained and tested; within each fold, the data are partitioned into training, validation, and test sets in a 6:2:2 ratio. The batch size is 32, and the learning rate for both the selector and the proxy-task model is set to $0.001$. For the SSVEP paradigm, the model is pre-trained for 500 epochs, followed by collaborative optimization for an additional 500 epochs. For collaborative optimization, AdamW with the AMSGrad variant is employed. The learning rate for both the selector and the proxy-task model is initialized at $0.001$ and follows cosine decay ($\alpha=0.9$). Gradients are clipped at a threshold of $1.0$. For the motor imagery/execution paradigm, pre-training and collaborative optimization are each performed for 50 epochs, since larger numbers of epochs are observed not to yield further improvement. The feature extraction backbone of the proposed framework is EEGNet~\cite{EEGNet2018}, unless otherwise specified.

\begin{figure*}[tb]
    \centering
      \includegraphics[width=7in]{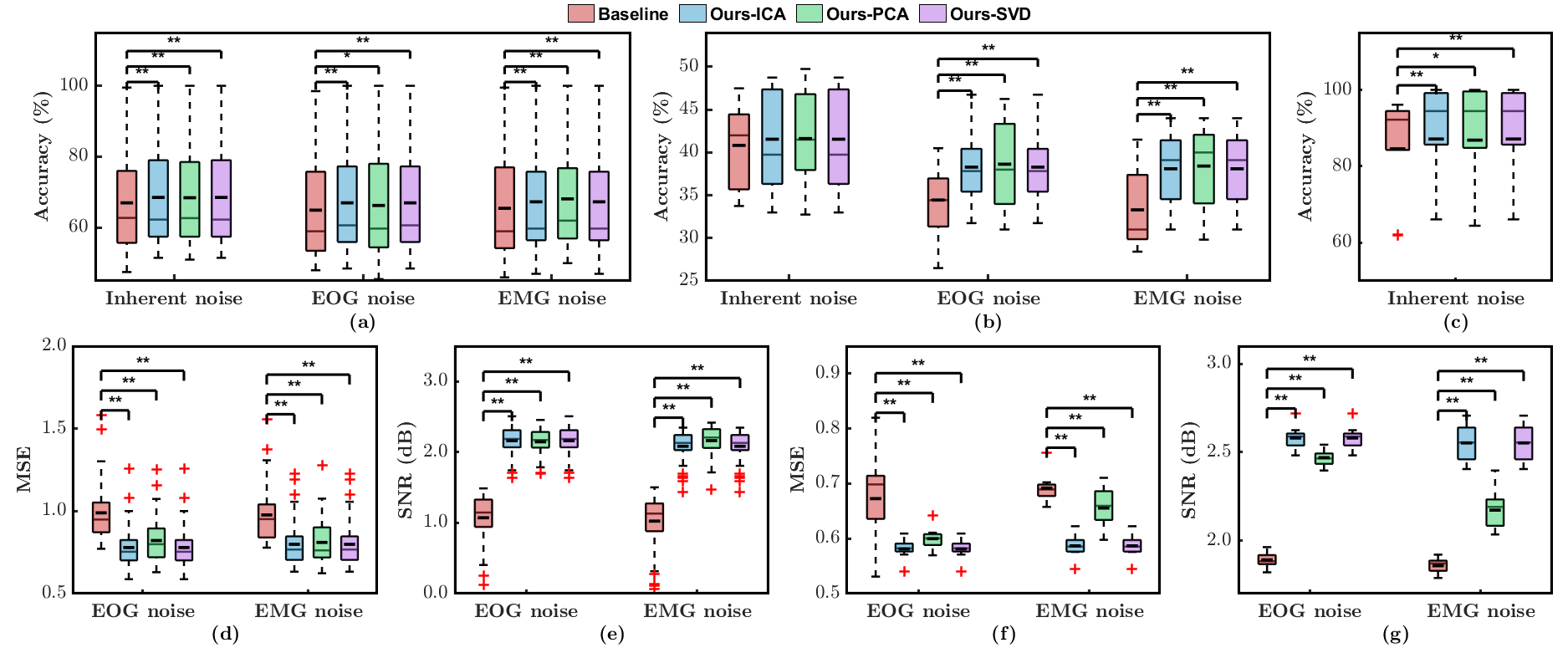}
        \caption{Comparison of EEG signals before denoising (Baseline) and after applying the proposed task-oriented learning framework with different decompositions (Ours–ICA/PCA/SVD). Task-related information: motor imagery~\cite{Cho2017} (a), motor execution~\cite{xiang2023learning} (b), SSVEP~\cite{nakanishi2015comparison} (c). Signal-quality metrics: motor imagery—MSE (d) and SNR (e); motor execution—MSE (f) and SNR (g). Paired Wilcoxon signed-rank tests compare pre- vs. post-denoising; * $p<0.05$, ** $p<0.01$.} 
     \centering
     \label{fig:Visual_result}
 \end{figure*}

\begin{table*}[!tb]\small
\begin{threeparttable}
\centering
\renewcommand{\arraystretch}{1.2}
\caption{Experimental results for denoising inherent instrumental and environmental noise on the motor imagery dataset~\cite{Cho2017}, the motor execution dataset~\cite{xiang2023learning}, and the SSVEP dataset~\cite{nakanishi2015comparison} (\%).}
\begin{tabular}{ p{1.5cm} p{0.95cm}<{\centering} p{0.95cm}<{\centering} p{0.95cm}<{\centering} p{0.95cm}<{\centering} p{0.95cm}<{\centering} p{0.95cm}<{\centering} p{0.95cm}<{\centering} p{0.95cm}<{\centering} p{0.95cm}<{\centering} p{0.95cm}<{\centering} p{0.95cm}<{\centering} p{0.95cm}<{\centering}}
\toprule
 \multirow{2}{*}{\textbf{Methods}} & \multicolumn{4}{c}{\textbf{Motor imagery dataset}} & \multicolumn{4}{c}{\textbf{Motor execution dataset}} & \multicolumn{4}{c}{\textbf{SSVEP dataset}}\\
& \textbf{Acc.} & \textbf{Prec.}  & \textbf{Recall} & \textbf{F1} & \textbf{Acc.} & \textbf{Prec.} & \textbf{Recall} & \textbf{F1} & \textbf{Acc.} & \textbf{Prec.} & \textbf{Recall} & \textbf{F1} \\
\midrule

Baseline & $66.99$ & $67.49$ & $67.20$ & $66.45$ & $40.83$ & $40.12$ & $40.71$ & $39.31$ & $84.56$ & $85.06$ & $84.95$ & $82.22$\\
\midrule
 ICA \cite{MNE-ICALabel2022} & $65.53$ & $66.04$ & $65.73$ & $64.97$ & $39.84$ & $39.77$ & $39.84$ & $38.32$ & $81.54$ & $81.65$ & $81.31$ & $79.52$\\
 PCA \cite{de2007denoising} & $67.54$ & $67.97$ & $67.60$ & $67.00$ & $40.88$ & $40.94$ & $40.91$ & $39.65$ & $78.58$ & $79.89$ & $79.46$ & $77.23$\\
 SVD \cite{gong2017improved} & $66.39$ & $67.08$ & $66.59$ & $65.61$ & $39.33$ & $39.34$ & $39.36$ & $38.01$ & $84.50$ & $84.87$ & $85.33$ & $83.36$\\
\midrule
Ours-ICA & $\mathbf{68.55}$ & $\mathbf{69.03}$ & $\mathbf{68.62}$ & $\mathbf{68.19}$ & $41.55$ & $41.04$ & $41.54$ & $40.32$ & $87.16$ & $88.40$ & $87.41$ & $86.32$\\
Ours-PCA & $68.45$ & $68.92$ & $68.56$ & $68.14$ & $41.62$ & $41.18$ & $41.49$ & $40.21$ & $86.85$ & $87.87$ & $87.44$ & $86.03$\\
Ours-SVD & $68.39$ & $68.80$ & $68.48$ & $68.09$ & $\mathbf{41.81}$ & $\mathbf{41.42}$ & $\mathbf{41.74}$ & $\mathbf{40.56}$ & $\mathbf{88.02}$ & $\mathbf{88.78}$ & $\mathbf{88.15}$ & $\mathbf{87.17}$\\

\bottomrule
      \end{tabular}

\begin{tablenotes}
    \item[1] Baseline denotes the raw EEG signal, which is inherently contaminated with instrumental and environmental noise.
\end{tablenotes}
\label{table: raw}
\end{threeparttable}
\end{table*}

\begin{table*}[!htb]\small
\begin{threeparttable}
\centering
\renewcommand{\arraystretch}{1.2}
\caption{Experimental results for denoising EOG noise on the motor imagery dataset~\cite{Cho2017} and the motor execution dataset~\cite{xiang2023learning} (\%).}
\begin{tabular}{p{2.4cm}
                p{2.4cm}<{\centering}
                p{2.4cm}<{\centering}
                p{1cm}<{\centering}
                p{1cm}<{\centering}
                p{1.2cm}<{\centering}
                p{1cm}<{\centering}
                p{1cm}<{\centering}
                p{1cm}<{\centering}}
\toprule
\multirow{2}{*}{\textbf{Noise}} 
& \multirow{2}{*}{\textbf{Model Type}} 
& \multirow{2}{*}{\textbf{Methods}} 
& \multicolumn{4}{c}{\textbf{Task-related Information}} 
& \multicolumn{2}{c}{\textbf{Signal Quality}} \\
& & & \textbf{Acc. $\uparrow$} & \textbf{Prec. $\uparrow$} & \textbf{Recall $\uparrow$} & \textbf{F1 $\uparrow$} & \textbf{MSE $\downarrow$} & \textbf{SNR $\uparrow$} \\
\midrule
\multirow{13}{*}{Motor imagery} 
&
Baseline & EEGNet \cite{EEGNet2018} & $64.94$ & $65.33$ & $65.05$ & $64.42$ & $0.98$ & $1.07$ \\
\cmidrule(l){2-9}
&
\multirow{3}{*}{Traditional} 
  & ICA \cite{MNE-ICALabel2022} & $63.17$ & $63.51$ & $63.35$ & $62.56$ & $1.07$ & $0.82$ \\
  && PCA \cite{de2007denoising} & $65.22$ & $65.69$ & $65.36$ & $64.65$ & $1.69$ & $1.50$ \\
  && SVD \cite{gong2017improved} & $64.44$ & $64.84$ & $64.66$ & $63.49$ & $0.78$ & $2.16$ \\
\cmidrule(l){2-9}
&
\multirow{3}{*}{Single-Channel (DL)} 
  & 1D-ResCNN \cite{1D-ResCNN2020} & $62.92$ & $63.51$ & $63.17$ & $62.15$ & $0.61$ & $3.20$ \\
  && DuoCL \cite{DuoCL2022} & $63.41$ & $63.90$ & $63.60$ & $62.64$ & $0.62$ & $3.24$ \\
  && EEGDNet \cite{EEGDNet2022} & $63.21$ & $63.99$ & $63.50$ & $62.20$ & $0.58$ & $\mathbf{3.72}$ \\
\cmidrule(l){2-9}
&
\multirow{3}{*}{Multi-Channels (DL)} 
  & EEGANet \cite{EEGANet2021} & $63.48$ & $63.96$ & $63.66$ & $62.36$ & $0.83$ & $3.22$ \\
  && STFNet \cite{STFNet2024} & $62.31$ & $62.99$ & $62.56$ & $61.13$ & $\mathbf{0.54}$ & $3.08$ \\
  && IC-U-Net \cite{IC-U-Net2022} & $59.70$ & $59.26$ & $60.21$ & $55.51$ & $1.06$ & $0.73$ \\
\cmidrule(l){2-9}
&
\multirow{3}{*}{Ours} 
  & Ours-ICA & $\mathbf{66.99}$ & $\mathbf{67.44}$ & $\mathbf{67.14}$ & $\mathbf{66.60}$ & $0.77$ & $2.16$ \\
  && Ours-PCA & $66.26$ & $66.64$ & $66.41$ & $65.89$ & $0.82$ & $2.14$ \\
  && Ours-SVD & $66.63$ & $67.07$ & $66.77$ & $66.26$ & $0.82$ & $2.14$ \\
\midrule
\multirow{13}{*}{Motor execution} 
&
Baseline & EEGNet \cite{EEGNet2018} & $34.44$ & $33.33$ & $33.69$ & $32.83$ & $0.67$ & $1.89$\\
\cmidrule(l){2-9}
&
\multirow{3}{*}{Traditional} 
& ICA \cite{MNE-ICALabel2022} & $34.73$ & $34.97$ & $34.63$ & $33.35$ & $0.73$ & $1.63$\\
&& PCA \cite{de2007denoising} & $33.75$ & $34.18$ & $33.96$ & $32.65$ & $1.23$ & $0.80$ \\
&& SVD \cite{gong2017improved} & $34.92$ & $34.32 $ & $34.63$ & $32.99$ & $0.60$ & $2.39$ \\
\cmidrule(l){2-9}
&
\multirow{3}{*}{Single-Channel (DL)} 
& 1D-ResCNN \cite{1D-ResCNN2020} & $31.38$ & $30.92$ & $30.65$ & $29.84$ & $0.50$ & $3.14$ \\
&& DuoCL \cite{DuoCL2022} & $35.18$ & $34.24$ & $34.55$ & $33.68$ & $0.67$ & $1.72$ \\
&& EEGDNet \cite{EEGDNet2022} & $34.55$ & $33.97$ & $33.64$ & $32.90$ & $\mathbf{0.42}$ & $4.09$ \\
\cmidrule(l){2-9}
&
\multirow{3}{*}{Multi-Channels (DL)} 
& EEGANet \cite{EEGANet2021} & $37.20$ & $36.60$ & $36.27$ & $35.43$ & $0.43$ & $\mathbf{4.22}$ \\
&& STFNet \cite{STFNet2024} & $36.66$ & $35.52$ & $35.61$ & $34.51$ & $0.43$ & $3.80$ \\
&& IC-U-Net \cite{IC-U-Net2022} & $36.97$ & $36.73$ & $36.15$ & $34.92$ & $0.64$ & $1.94$ \\
\cmidrule(l){2-9}
&
\multirow{3}{*}{Ours} 
  & Ours-ICA & $38.29$ & $37.66$ & $38.14$ & $36.97$ & $0.58$ & $2.58$ \\
  && Ours-PCA & $38.63$ & $38.80$ & $38.65$ & $37.65$ & $0.60$ & $2.47$ \\
  && Ours-SVD & $\mathbf{38.98}$ & $\mathbf{38.84}$ & $\mathbf{39.25}$ & $\mathbf{38.00}$ & $0.60$ & $2.47$ \\
\bottomrule
\end{tabular}
\begin{tablenotes}
    \item[1] $\uparrow$: higher is better; $\downarrow$: lower is better. 
    \item[2] Baseline denotes the raw EEG signal added with EOG noise.
    \item[3] DL denotes deep-learning-based algorithms.

\end{tablenotes}
\label{table: EOG}
\end{threeparttable}
\end{table*}

\begin{table*}[!htb]\small
\begin{threeparttable}
\centering
\renewcommand{\arraystretch}{1.2}
\caption{Experimental results for denoising EMG noise on the motor imagery dataset~\cite{Cho2017} and the motor execution dataset~\cite{xiang2023learning} (\%).}
\begin{tabular}{p{2.4cm}
                p{2.4cm}<{\centering}
                p{2.4cm}<{\centering}
                p{1cm}<{\centering}
                p{1cm}<{\centering}
                p{1.2cm}<{\centering}
                p{1cm}<{\centering}
                p{1cm}<{\centering}
                p{1cm}<{\centering}}
\toprule
\multirow{2}{*}{\textbf{Noise}} 
& \multirow{2}{*}{\textbf{Model Type}} 
& \multirow{2}{*}{\textbf{Methods}} 
& \multicolumn{4}{c}{\textbf{Task-related Information}} 
& \multicolumn{2}{c}{\textbf{Signal Quality}} \\
& & & \textbf{Acc. $\uparrow$} & \textbf{Prec. $\uparrow$} & \textbf{Recall $\uparrow$} & \textbf{F1 $\uparrow$} & \textbf{MSE $\downarrow$} & \textbf{SNR $\uparrow$} \\
\midrule

\multirow{13}{*}{Motor imagery} 
&
\multirow{1}{*}{Baseline} 
& EEGNet \cite{EEGNet2018} & $65.47$ & $66.02$ & $65.62$ & $64.88$ & $0.97$ & $1.02$\\
\cmidrule(l){2-9}
&
\multirow{3}{*}{Traditional} 
& ICA \cite{MNE-ICALabel2022} & $65.33$ & $65.80$ & $65.46$ & $64.70$ & $1.02$ & $0.86$\\
&& PCA \cite{de2007denoising} & $66.65$ & $67.00$ & $66.76$ & $66.10$ & $1.36$ & $0.30$\\
&& SVD \cite{gong2017improved} & $65.81$ & $66.49$ & $66.01$ & $64.91$ & $0.78$ & $2.12$\\

\cmidrule(l){2-9}
&
\multirow{3}{*}{Single-Channel (DL)}  
& 1D-ResCNN \cite{1D-ResCNN2020} & $63.92$ & $64.43$ & $64.15$ & $63.14$ & $0.56$ & $3.42$ \\
&& DuoCL \cite{DuoCL2022} & $64.70$ & $65.43$ & $64.86$ & $64.04$ & $0.49$ & $ 4.27$ \\
&& EEGDNet \cite{EEGDNet2022} & $65.16$ & $65.92$ & $65.35$ & $64.47$ & $0.46$ & $\mathbf{4.72}$ \\
\cmidrule(l){2-9}
&
\multirow{3}{*}{Multi-Channels (DL)} 
& EEGANet \cite{EEGANet2021} & $64.64$ & $65.30$ & $64.76$ & $63.48$ & $0.77$ & $4.02$ \\
&& STFNet \cite{STFNet2024} & $63.38$ & $63.75$ & $63.55$ & $62.43$ & $\mathbf{0.38}$ & $3.53$ \\
&& IC-U-Net \cite{IC-U-Net2022} & $60.01$ & $59.99$ & $60.39$ & $56.49$ & $1.04$ & $0.93$ \\
\cmidrule(l){2-9}
&
\multirow{3}{*}{Ours} 
  & Ours-ICA & $67.31$ & $67.84$ & $67.43$ & $66.92$ & $0.80$ & $2.08$ \\
  && Ours-PCA & $\mathbf{68.11}$ & $\mathbf{68.69}$ & $\mathbf{68.26}$ & $\mathbf{67.72}$ & $0.81$ & $2.16$ \\
  && Ours-SVD & $67.87$ & $68.45$ & $67.99$ & $67.46$ & $0.82$ & $2.15$ \\
\midrule

\multirow{13}{*}{Motor execution} 
&
\multirow{1}{*}{Baseline} 
& EEGNet \cite{EEGNet2018} & $33.29$ & $32.08$ & $33.27$ & $31.81$ & $0.69$ & $1.85$\\
\cmidrule(l){2-9}
&
\multirow{3}{*}{Traditional} 
& ICA \cite{MNE-ICALabel2022} & $34.11$ & $34.08$ & $33.95$ & $32.78$ & $0.72$ & $1.72$\\
&& PCA \cite{de2007denoising} & $34.14$ & $34.12$ & $34.09$ & $32.71$ & $1.23$ & $0.79$ \\
&& SVD \cite{gong2017improved} & $35.91$ & $35.91$ & $35.77$ & $34.10$ & $0.60$ & $2.41$ \\

\cmidrule(l){2-9}
&
\multirow{3}{*}{Single-Channel (DL)}  
& 1D-ResCNN \cite{1D-ResCNN2020} & $31.70$ & $31.63$ & $31.64$ & $29.84$ & $0.50$ & $3.14$ \\
&& DuoCL \cite{DuoCL2022} & $35.49$ & $35.11$ & $35.27$ & $34.02$ & $0.67$ & $1.85$ \\
&& EEGDNet \cite{EEGDNet2022} & $34.12$ & $33.97$ & $33.64$ & $32.90$ & $\mathbf{0.42}$ & $4.09$ \\
\cmidrule(l){2-9}
&
\multirow{3}{*}{Multi-Channels (DL)} 
& EEGANet \cite{EEGANet2021} & $36.21$ & $36.17$ & $36.08$ & $34.65$ & $0.43$ & $\mathbf{4.19}$ \\
&& STFNet \cite{STFNet2024} & $36.76$ & $36.21$ & $36.65$ & $34.96$ & $0.44$ & $3.83$ \\
&& IC-U-Net \cite{IC-U-Net2022} & $35.56$ & $35.02$ & $35.60$ & $33.54$ & $0.64$ & $2.01$ \\
\cmidrule(l){2-9}
&
\multirow{3}{*}{Ours} 
  & Ours-ICA & $38.09$ & $37.98$ & $38.19$ & $37.07$ & $0.58$ & $2.55$ \\
  && Ours-PCA & $\mathbf{38.41}$ & $\mathbf{38.42}$ & $\mathbf{38.37}$ & $\mathbf{37.42}$ & $0.65$ & $2.17$ \\
  && Ours-SVD & $37.27$ & $37.02$ & $37.20$ & $36.06$ & $0.65$ & $2.19$ \\
\bottomrule
\end{tabular}
\begin{tablenotes}
    \item[1] $\uparrow$: higher is better; $\downarrow$: lower is better. 
    \item[2] Baseline denotes the raw EEG signal added with EMG noise.
    \item[3] DL denotes deep-learning-based algorithms.
\end{tablenotes}
\label{table: EMG}
\end{threeparttable}

\end{table*}

\subsection{Experimental Results}

\subsubsection{Denoising Performance Across Paradigms}

Experiments are conducted on three datasets covering distinct tasks, which are grouped into two paradigms: SSVEP and motor imagery/execution. The task-related information differs between these paradigms: SSVEP primarily relies on frequency-domain features, whereas motor imagery/execution depends on combined frequency–spatial–temporal representations. This setting tests whether the task-oriented learning framework accommodates various signal characteristics.

Results (Tables~\ref{table: raw}, \ref{table: EOG}, and \ref{table: EMG}) indicate that the task-oriented learning framework generalizes across paradigms. Because ocular and muscular artifacts typically have limited impact on signals recorded over the occipital cortex in SSVEP, only inherent noise is considered for that paradigm; improvements over the baseline raw signals are observed (Acc.: $2.78\%\uparrow$, Prec.: $3.29\%\uparrow$, Recall: $2.72\%\uparrow$, F1: $4.29\%\uparrow$). For motor imagery/execution, average gains across inherent, EOG, and EMG noise conditions are as follows (motor imagery/motor execution): Acc.: $1.82\%/3.22\%\uparrow$, Prec.: $1.82\%/3.97\%\uparrow$, Recall: $1.78\%/3.51\%\uparrow$, F1: $2.00\%/3.60\%\uparrow$. Signal-quality metrics under physiological-noise conditions (motor imagery/motor execution) also improve: MSE decreases by $0.17/0.07$, while SNR increases by $1.10/0.54$\,dB. These findings suggest that the selector, guided by task labels, identifies informative components from the decomposed EEG regardless of the properties of EEG characteristics. It supports the task-oriented learning framework’s potential as a general-purpose denoising solution for EEG signals.

\subsubsection{Denoising Performance Across Noises}

EEG signals are contaminated by multiple noise sources. The proposed task-oriented learning framework is evaluated on both simulated physiological artifacts (EOG, EMG) and inherent instrumental and environmental noise, and consistently shows improvements in signal quality under these contamination conditions.

For inherent noise, averaged over three datasets, task-related metrics improve relative to the baseline: Acc.: $1.69\%\uparrow$, Prec.: $1.94\%\uparrow$, Recall: $1.65\%\uparrow$, F1: $2.34\%\uparrow$. For simulated physiological artifacts, evaluated on the motor imagery and motor execution datasets, gains are observed in both task-related metrics and signal-quality measures. Task-related metrics (EOG/EMG) improve as follows: Acc.: $2.94\%/3.46\%\uparrow$, Prec.: $3.41\%/4.02\%\uparrow$, Recall: $3.36\%/3.46\%\uparrow$, F1: $3.27\%/3.76\%\uparrow$. Signal-quality metrics (EOG/EMG) also improve: MSE decreases by $0.13/0.11$, while SNR increases by $0.85/0.78$\,dB. These results indicate that task-related label guidance enables the selector to identify informative components and suppress noise across diverse conditions.

Figure~\ref{fig:Visual_result} visualizes subject-level distributions for baseline versus processed signals across datasets. All metrics show improvement on average. Moreover, except for accuracy under inherent noise in the motor execution dataset, all other comparisons show statistically significant improvements according to the paired Wilcoxon signed rank tests ($p<0.05$, and $p<0.01$ in most cases), further supporting the effectiveness of the task-oriented learning framework.

\subsubsection{Competitive Performance for EEG Denoising}

To comprehensively assess performance, the proposed task-oriented learning framework is compared against representative EEG denoising methods spanning traditional signal-processing–based~\cite{MNE-ICALabel2022, de2007denoising, gong2017improved} and learning-based approaches~\cite{1D-ResCNN2020, DuoCL2022, EEGDNet2022, EEGANet2021, STFNet2024, IC-U-Net2022}. Traditional BSS-based methods are included, and the task-oriented learning framework is also evaluated in combination with the same BSS methods. The learning-based baselines cover CNN-~\cite{1D-ResCNN2020, IC-U-Net2022}, RNN-~\cite{DuoCL2022}, and transformer-based autoencoders~\cite{EEGDNet2022, STFNet2024}, as well as GANs~\cite{EEGANet2021}. They further span two design routes: single-channel denoising (processing each channel independently)~\cite{1D-ResCNN2020, DuoCL2022, EEGDNet2022} and multi-channel denoising (jointly processing all channels)~\cite{EEGANet2021, STFNet2024, IC-U-Net2022}.

Deep-learning–based methods achieve strong performance on signal-quality metrics (MSE, SNR), attaining the best scores in Tables~\ref{table: EOG} and \ref{table: EMG}. This outcome is expected because training leverages clean reference signals and employs objectives, such as minimizing MSE or matching the distribution of clean signals, directly aligned with these metrics. However, task-relevant information can be degraded, as reflected by reduced performance on proxy downstream classification tasks. For example, on the motor imagery dataset with EOG/EMG artifacts~\cite{Cho2017}, signals denoised by learning-based baselines yield lower downstream performance than the original noisy inputs. These observations indicate that gains in MSE/SNR do not necessarily translate into improved utility for downstream inference, which is the key motivation for denoising.

Moreover, because truly clean EEG references are unavailable for the inherent instrumental and environmental noises, supervised learning methods that require clean targets have been excluded from the comparison in Table~\ref{table: raw}. This underscores a key advantage of the proposed task-oriented learning framework: training relies solely on readily available proxy-task labels, enabling applicability across diverse noise sources, including those inherent to EEG.

Traditional signal-processing–based methods often underperform on reconstruction metrics yet cause limited degradation in downstream accuracy. This reflects a strength (minimal artifact introduction through manual component selection) but also a weakness (the difficulty of reliably selecting informative components). The proposed task-oriented learning framework leverages decomposition robustness while mitigating the selection challenge via a selector trained under task-related label supervision, yielding improvements in both signal quality and preservation of task-relevant information.

In summary, although the task-oriented learning framework yields weaker reconstruction metrics (e.g., MSE and SNR) than deep-learning methods trained with clean EEG references, it does not require clean signals for training, which enhances its applicability in real-world scenarios. Furthermore, supervision via task labels promotes retention of task-relevant information, producing denoised signals that are more amenable to downstream analysis.

\subsubsection{Retention of task-related information during denoising}

The motivation for EEG denoising is to facilitate analysis of task-related information. Therefore, beyond signal-quality metrics such as SNR and MSE, a critical criterion is whether task-relevant content is preserved through the denoising process. Because task-related components are often weak and entangled with noise, reconstruction metrics may not fully capture their integrity.

A proxy-task classifier is employed to quantify task-relevant information in the denoised EEG. As summarized in Tables~\ref{table: raw}, \ref{table: EOG}, and \ref{table: EMG}, the proposed task-oriented learning framework exhibits consistent advantages. Considering accuracy (averaged over the compared methods), the gains are:
\begin{itemize}
  \item \textbf{SSVEP}: $2.78\%/2.84\% \uparrow$ (inherent noise).
  \item \textbf{Motor imagery}: $1.47\%/0.92\% \uparrow$ (inherent noise), $1.69\%/1.41\% \uparrow$ (EOG noise), $2.29\%/1.11\% \uparrow$ (EMG noise).
  \item \textbf{Motor execution}: $0.83\%/0.78\% \uparrow$ (inherent noise), $4.19\%/1.43\% \uparrow$ (EOG noise), $4.63\%/1.16\% \uparrow$ (EMG noise).
\end{itemize}

Here, each pair denotes improvement over the baseline / over the best competing method. Note that the strongest non-proposed method differs across settings; nevertheless, the proposed pipeline delivers the most consistently superior accuracy, indicating that the denoising process preserves and makes more salient task-related information.

\begin{table}[!tb]\small
\begin{threeparttable}
\centering
\renewcommand{\arraystretch}{1.2}
\caption{Comparison of models trained with the learned selector versus uniform random component mixing on the SSVEP dataset~\cite{nakanishi2015comparison} (\%).}
\begin{tabular}{p{1.3cm}
                p{1.3cm}<{\centering}
                p{0.9cm}<{\centering}
                p{0.9cm}<{\centering}
                p{1.0cm}<{\centering}
                p{0.9cm}<{\centering}}

\toprule
\textbf{Algorithm} & \textbf{Mode} & \textbf{Acc.}& \textbf{Prec.} & \textbf{Recall}& \textbf{F1}  \\
\cmidrule{1-6}
  \multicolumn{2}{c}{Baseline} & $84.56$ & $85.06$ & $84.95$ & $82.22$ \\
\cmidrule{1-6}
\multirow{2}{*}{ICA} & Random & $81.29$ & $81.68$ & $81.47$ & $79.20$ \\
  & Selector & $\mathbf{87.16}$ & $\mathbf{88.40}$ & $\mathbf{87.41}$ & $\mathbf{86.32}$ \\
\cmidrule{1-6}
\multirow{2}{*}{PCA} & Random& $82.40$ & $82.48$ & $83.33$ & $80.59$ \\
  & Selector & $\mathbf{86.85}$ & $\mathbf{87.87}$ & $\mathbf{87.44}$ & $\mathbf{86.03}$ \\
\cmidrule{1-6}
\multirow{2}{*}{SVD} & Random & $83.58$ & $83.57$ & $84.00$ & $81.95$ \\
  & Selector & $\mathbf{88.02}$ & $\mathbf{86.78}$ & $\mathbf{88.15}$ & $\mathbf{87.17}$ \\
\bottomrule
\end{tabular}
\begin{tablenotes}
    \item[1] Baseline denotes the raw EEG signal, which is inherently contaminated with instrumental and environmental noise.
\end{tablenotes}
\label{table:result:Nakanishi2015}
\end{threeparttable}
\end{table}

\begin{table}[!tb]\small
\centering
\renewcommand{\arraystretch}{1.2}
\caption{Comparison of collaborative optimization with stabilization techniques (Ours) versus the baseline without (BL) (\%).
}
\begin{threeparttable}
\begin{tabular}{p{1cm}
                p{0.8cm}<{\centering}
                p{0.8cm}<{\centering}
                p{0.8cm}<{\centering}
                p{0.8cm}<{\centering}
                p{0.8cm}<{\centering}
                p{0.8cm}<{\centering}}

\toprule
\textbf{Noise} & \textbf{Dataset} & \textbf{Mode} & \textbf{Acc.}& \textbf{Prec.} & \textbf{Recall}& \textbf{F1}  \\
\cmidrule{1-7}
\multirow{6}{*}{Inherent} & \multirow{2}{*}{MI} & BL & $68.19$ & $68.58$ & $68.26$ & $67.86$ \\
  & & Ours  & $\mathbf{68.46}$ & $\mathbf{68.92}$ & $\mathbf{68.55}$ & $\mathbf{68.14}$ \\
  \cmidrule{2-7}
& \multirow{2}{*}{ME} & BL & $\textbf{41.81}$ & $\textbf{41.69}$ & $\textbf{41.89}$ & $\textbf{40.65}$ \\
  & & Ours  & $41.66$ & $41.21$ & $41.59$ & $40.36$ \\
  \cmidrule{2-7}
& \multirow{2}{*}{SSVEP} & BL & $86.11$ & $86.15$ & $86.51$ & $84.56$ \\
  & & Ours  & $\mathbf{87.34}$ & $\mathbf{88.35}$ & $\mathbf{87.67}$ & $\mathbf{86.51}$ \\
\cmidrule{1-7}
\multirow{4}{*}{EOG} & \multirow{2}{*}{MI} & BL & $\textbf{66.71}$ & $\textbf{67.10}$ & $\textbf{66.82}$ & $\textbf{66.36}$ \\
  & & Ours  & $66.63$ & $67.05$ & $66.77$ & $66.25$ \\
  \cmidrule{2-7}
& \multirow{2}{*}{ME} & BL & $38.02$ & $37.67$ & $37.71$ & $36.74$ \\
  & & Ours  & $\textbf{38.63}$ & $\textbf{38.43}$ & $\textbf{38.68}$ & $\textbf{37.54}$\\
\cmidrule{1-7}
\multirow{4}{*}{EMG} & \multirow{2}{*}{MI} & BL & $67.58$ & $67.97$ & $67.68$ & $67.24$ \\
  & & Ours  & $\textbf{67.76}$ & $\textbf{68.33}$ & $\textbf{67.89}$ & $\textbf{67.37}$ \\
  \cmidrule{2-7}
& \multirow{2}{*}{ME} & BL & $36.44$ & $36.28$ & $36.45$ & $35.31$ \\
  & & Ours  & $\textbf{37.92}$ & $\textbf{37.81}$ & $\textbf{37.92}$ & $\textbf{36.85}$ \\
\bottomrule
\end{tabular}
\begin{tablenotes}
    \item[1] Results in this table are averaged across decomposition methods (PCA, ICA, SVD).
    \item[2] MI: motor imagery; ME: motor execution.
\end{tablenotes}
\label{table:result:ablation}
\end{threeparttable}
\end{table}

\subsection{Ablation Study}

\subsubsection{The Component Selection Strategy of Selector}

The probabilistic outputs of the selector change during the training, and thus introduce diversity in reconstructed EEG signals, which could act as data augmentation for training decoders. This raises the question of whether observed improvements in decoder performance stem from increased data diversity or from enhanced signal quality due to denoising. To disentangle these effects, an ablation study is conducted on the SSVEP dataset~\cite{nakanishi2015comparison}: comparison decoders are trained on EEG reconstructed by combining decomposed components with coefficients sampled uniformly at random. Three decomposition algorithms are evaluated within the proposed task-oriented learning framework.

As shown in Table~\ref{table:result:Nakanishi2015}, relative to the baseline trained on raw EEG, the random-mixing control degrades performance (Acc.: $2.14\%\downarrow$, Prec.: $2.48\%\downarrow$, Recall: $2.02\%\downarrow$, F1: $1.64\%\downarrow$). In contrast, reconstruction guided by the learned selector improves performance (Acc.: $2.78\%\uparrow$, Prec.: $3.29\%\uparrow$, Recall: $2.72\%\uparrow$, F1: $4.29\%\uparrow$). These findings indicate that naively varying component ratios does not constitute an effective data-augmentation strategy for EEG decoders; the gains arise from improved EEG quality rather than diversity alone.

\subsubsection{Stabilization Techniques for Collaborative Optimization}

To help satisfy the hypotheses (\textbf{H1}–\textbf{H4}) in Sec.~\ref{Sec. Convergence Analysis}, several optimization techniques are employed, including gradient clipping and AdamW with the AMSGrad variant. Although these techniques do not guarantee that the hypotheses hold, they encourage the optimization dynamics to conform to the stated conditions. To assess their contribution, an ablation compares training with and without these techniques. For robustness, the results in Table~\ref{table:result:ablation} are averaged across three decomposition methods (ICA, PCA, SVD).

Across five of seven evaluation conditions, the stabilization techniques improve performance relative to training without them. Furthermore, the gains (Acc.: $0.75\%\uparrow$; Prec.: $1.04\%\uparrow$; Recall: $0.82\%\uparrow$; F1: $0.94\%\uparrow$) exceed the occasional declines in magnitude (Acc.: $0.11\%\downarrow$; Prec.: $0.27\%\downarrow$; Recall: $0.18\%\downarrow$; F1: $0.20\%\downarrow$). These findings indicate small but consistent benefits for collaborative training. The modest overall effect likely reflects the task-oriented learning framework's relative simplicity and its intrinsic adherence to the stated assumptions, which limit the potential for further improvement from the optimization.

\section{Discussion}

\begin{figure*}[tb]
    \centering
      \includegraphics[width=7in]{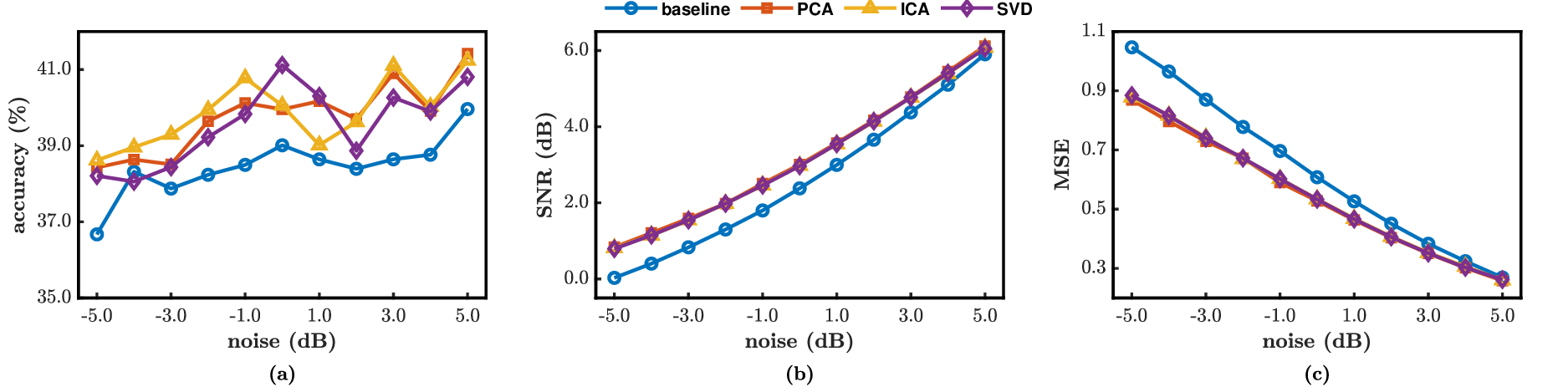}
        \caption{Performances on the motor-execution dataset \cite{xiang2023learning} across noise levels.}
     \centering
     \label{fig:noise_SNR}
 \end{figure*}

\begin{table*}[!tb]\small
\begin{threeparttable}

\centering
\renewcommand{\arraystretch}{1.2}
\caption{Experimental results for the proposed task-oriented learning framework combined with different learning-based feature extractors on the SSVEP dataset~\cite{nakanishi2015comparison} (\%).}
\begin{tabular}{ p{1.5cm} p{0.95cm}<{\centering} p{0.95cm}<{\centering} p{0.95cm}<{\centering} p{0.95cm}<{\centering} p{0.95cm}<{\centering} p{0.95cm}<{\centering} p{0.95cm}<{\centering} p{0.95cm}<{\centering} p{0.95cm}<{\centering} p{0.95cm}<{\centering} p{0.95cm}<{\centering} p{0.95cm}<{\centering}}
\toprule
 \multirow{2}{*}{\textbf{Methods}} & \multicolumn{4}{c}{\textbf{EEGNeX~\cite{EEGNeX2024}}} & \multicolumn{4}{c}{\textbf{EEGTCNet~\cite{EEGTCNet2020}}} & \multicolumn{4}{c}{\textbf{DeepConvNet~\cite{DeepConvNet2017}}}\\
& \textbf{Acc.} & \textbf{Prec.}  & \textbf{Recall} & \textbf{F1} & \textbf{Acc.} & \textbf{Prec.} & \textbf{Recall} & \textbf{F1} & \textbf{Acc.} & \textbf{Prec.} & \textbf{Recall} & \textbf{F1} \\
\midrule

Baseline & $89.93$ & $90.45$ & $90.13$ & $89.13$ & $70.37$ & $70.93$ & $70.84$ & $68.95$ & $81.11$ & $81.72$ & $80.97$ & $79.10$\\

\midrule
 ICA \cite{MNE-ICALabel2022} & $85.80$ & $85.52$ & $85.62$ & $84.05$ & $64.75$ & $64.72$ & $64.84$ & $62.76$ & $71.35$ & $72.29$ & $71.89$ & $69.64$\\
 PCA \cite{de2007denoising} & $85.74$ & $86.56$ & $86.43$ & $85.30$ & $62.40$ & $63.23$ & $62.30$ & $60.08$ & $65.80$ & $67.04$ & $65.47$ & $63.33$\\
 SVD \cite{gong2017improved} & $85.24$ & $85.24$ & $86.56$ & $86.15$ & $69.38$ & $69.35$ & $69.87$ & $67.36$ & $78.14$ & $80.47$ & $78.55$ & $77.05$\\

\midrule

Ours-ICA & $\mathbf{92.90}$ & $\mathbf{92.82}$ & $\mathbf{92.70}$ & $\mathbf{92.20}$ & $\mathbf{77.34}$ & $\mathbf{76.87}$ & $\mathbf{77.18}$ & $\mathbf{75.42}$ & $92.71$ & $93.35$ & $92.95$ & $92.20$\\
Ours-PCA & $92.59$ & $92.80$ & $92.40$ & $91.99$ & $76.60$ & $76.16$ & $76.39$ & $74.68$ & $92.83$ & $\mathbf{94.04}$ & $93.23$ & $\mathbf{92.63}$\\
Ours-SVD & $92.65$ & $92.74$ & $92.55$ & $92.04$ & $76.72$ & $76.59$ & $76.52$ & $74.83$ & $\mathbf{93.14}$ & $93.66$ & $\mathbf{93.34}$ & $92.59$\\
\bottomrule
      \end{tabular}
\label{table: result: SSVEP}

\begin{tablenotes}
    \item[1] Baseline denotes the raw EEG signal, which is inherently contaminated with instrumental and environmental noise.
\end{tablenotes}
\end{threeparttable}

\end{table*}

\subsection{Performance Under Varying Noise Levels}

To assess robustness to different noise levels, an additional experiment was conducted on a motor execution dataset \cite{xiang2023learning} with synthetic EOG artifacts added at SNR ranging from $-5$ to $5$\,dB. Across all SNRs, both signal-quality metrics and task-related metrics of the denoised signals improved relative to the noisy baseline. The gains in signal-quality metrics are larger at more severe noise (e.g., near $-5$\,dB) than at milder noise (e.g., near $5$\,dB), whereas improvements in classification accuracy are comparatively uniform across SNRs. This pattern suggests that the task-guided selector can suppress not only the synthetic EOG artifacts but also inherent noise that is not explicitly simulated, thereby enhancing class separability. Such improvements may not always be fully captured by generic signal-quality metrics and can, in some cases, trade off with them.

\subsection{The Framework is Algorithm-agnostic}

The proposed task-oriented learning framework involves two parts: (1) the BSS-based decomposition module and (2) the selector and proxy-task models. To validate that the proposed framework is algorithm-agnostic, various decomposition algorithms and learning-based models are evaluated within the framework~\cite{EEGNet2018, EEGNeX2024, EEGTCNet2020, DeepConvNet2017}. The decomposition algorithms can be directly applied within the framework, while the learning-based models keep the representation learning module, changing the dimension of the output head for the designed tasks. 

For the BSS-based decomposition module, three widely used decomposition algorithms are evaluated across different tasks. The results obtained with these various decompositions are consistent and outperform their counterparts in traditional denoising pipelines, indicating compatibility of the proposed task-oriented learning framework with different decomposition algorithms.

For the selector and proxy-task models, an additional study is conducted on the SSVEP dataset~\cite{nakanishi2015comparison} using several network architectures~\cite{EEGNet2018, EEGNeX2024, EEGTCNet2020, DeepConvNet2017}. Although absolute performance varies across architectures, all demonstrate improved downstream classification compared to the original noisy EEG signals, implying that noisy components are being suppressed while task-related components are emphasized. This behavior aligns with the design motivation for denoising and further confirms that the task-oriented learning framework operates effectively in an algorithm-agnostic manner.

\begin{figure*}[tb]
    \centering
      \includegraphics[width=7in]{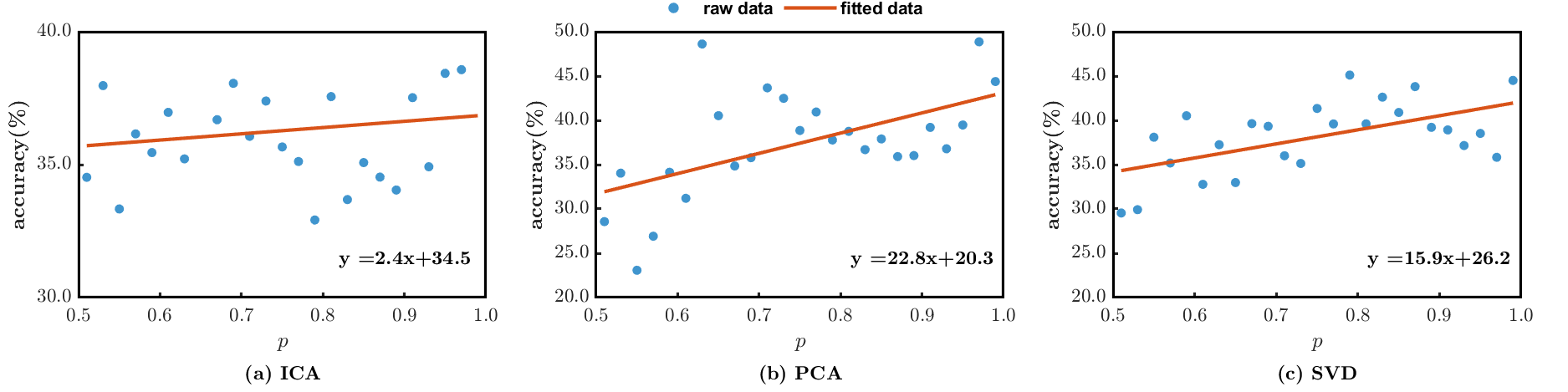}
        \caption{Component-level classification accuracy on the SSVEP dataset~\cite{nakanishi2015comparison} across BSS algorithms.}
     \centering
     \label{fig:component_acc}
 \end{figure*}

\begin{figure}[tb]
    \centering
      \includegraphics[width=3.5in]{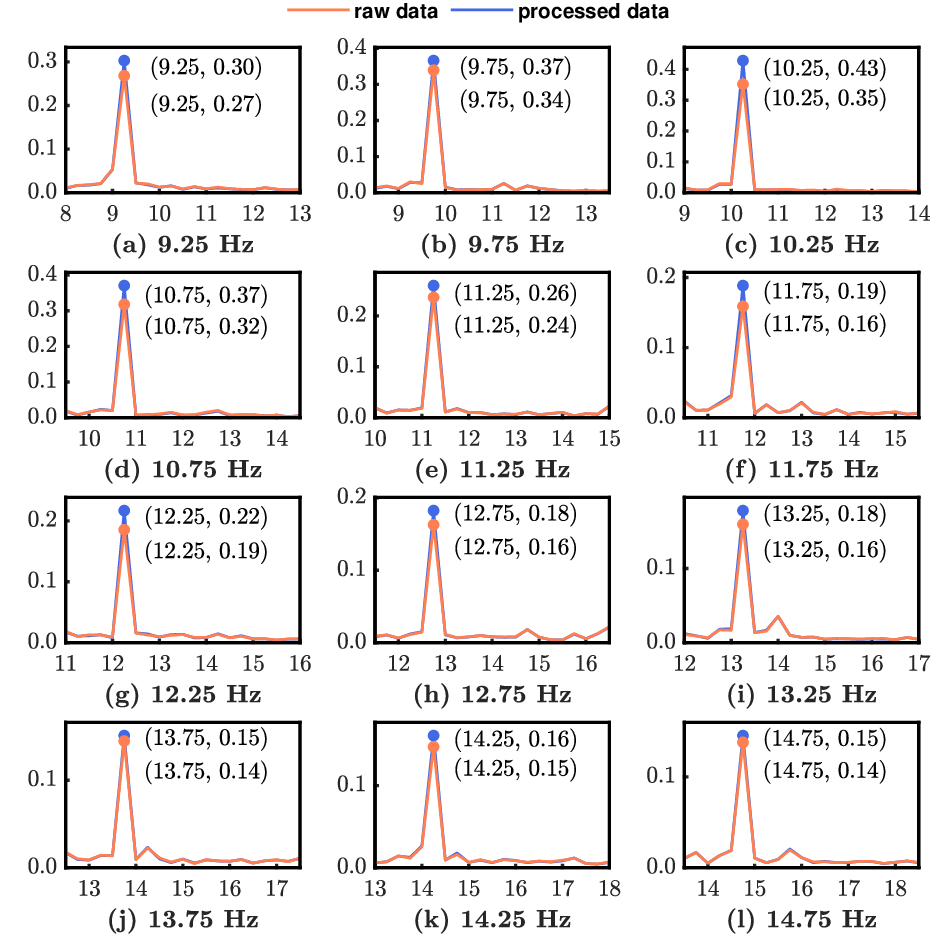}
        \caption{Comparison of power spectra and per-frequency power ratios for raw EEG and EEG processed by the proposed task-oriented learning framework using ICA-based decomposition for a representative subject from the SSVEP dataset~\cite{nakanishi2015comparison}. Panels (a)–(l) correspond to stimulus fundamental frequencies (9.25–14.75\,Hz in 0.5\,Hz steps); markers denote the peak magnitude at each target frequency.}
     \centering
     \label{fig:SSVEP}
 \end{figure}

 \begin{figure}[tb]
    \centering
      \includegraphics[width=3.5in]{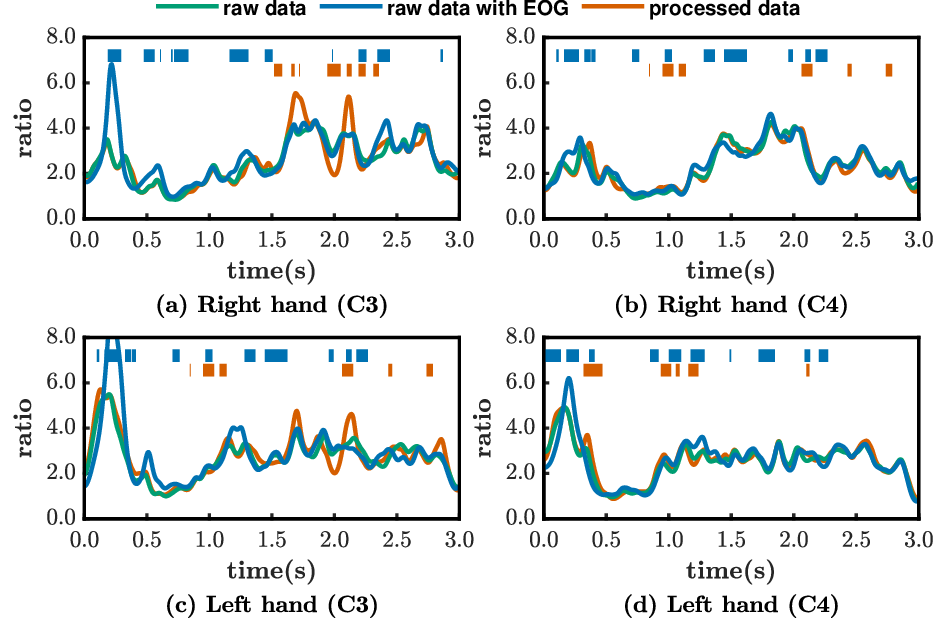}
        \caption{Alpha–to–beta (8–12\,Hz / 12–30\,Hz) power-ratio trajectories for raw EEG , raw EEG with added EOG, and EEG processed by the proposed task-oriented learning framework using ICA-based decomposition for a representative subject from the motor imagery dataset~\cite{Cho2017}. Colored significance bars along the top margin indicate time points significantly different from the raw baseline (paired two-tailed $t$-test): blue bars (upper track) for raw+EOG vs.\ raw, and orange bars (lower track) for processed vs.\ raw ($p<0.05$).}
     \centering
     \label{fig:MI}
 \end{figure}
 
\subsection{The Impact of the Selector in the Framework}

\subsubsection{Selector Favors Task-related Components}

The framework is designed to select task-related components from the decomposition, which are expected to be more identifiable by proxy-tasks. To assess this hypothesis, a classifier is trained at the component level. As shown in Fig.~\ref{fig:component_acc}, classification accuracy increases as the selector's retention probability $p$ (i.e., the probability assigned to keep a component) increases. This trend indicates that components favored by the selector are more informative for the proxy-task, supporting the idea that the selector prioritizes task-related components over noise. 

\subsubsection{Processed Data Exhibits Higher Fundamental-Frequency Power in the SSVEP Paradigm}

To examine why the denoised signals are easier to classify, the power spectrum and per-frequency power ratios are visualized. Relative to the raw recordings, the processed EEG exhibits a higher power ratio at the stimulus fundamental frequency (Fig.~\ref{fig:SSVEP}). This enhancement is crucial for SSVEP target identification, indicating that denoising preserves task-relevant oscillatory components while suppressing irrelevant activity. Consequently, the denoised signals present more prominent features for classification.

\subsubsection{Processed Data Exhibits More Coherent Spectrotemporal Features in the Motor Imagery Paradigm}

In motor imagery, the dominant EEG characteristic is the modulation of the sensorimotor rhythm (SMR), which is observed as event-related desynchronization/synchronization (ERD/ERS). Baseline variability introduced by EOG contamination complicates direct comparison of ERD/ERS waveforms between reference and denoised signals. Therefore, the alpha–to–beta power ratio is employed as a summary measure of SMR modulation, focusing on the alpha (8-12\,Hz) and beta (12-30\,Hz) bands most implicated in ERD/ERS.

A case study on the public dataset~\cite{Cho2017} (Fig.~\ref{fig:MI}) visualizes the alpha–to–beta power ratio under three conditions—raw data, raw data with added EOG, and the processed (denoised) signal—averaged over twenty imagery trials for channels C3 and C4, which are closely associated with motor cortex activity. Relative to the raw baseline, EOG contamination distorts the ratio trajectory, whereas the processed EEG recovers a trajectory that closely follows the raw pattern.

A time-point–wise $t$-test indicates that EOG-contaminated signals exhibit substantially more significant deviations from the raw data than the processed signals. This finding suggests that the proposed denoising procedure restores the underlying power distribution by suppressing artifact-related components while preserving task-relevant SMR modulation.

\section{Conclusions}

This study presents a task-oriented learning framework for automatic EEG denoising that uses task labels without clean reference EEG signals. The procedure decomposes the EEG into components using traditional BSS methods, applies a learning-based selector to assign retention probabilities, and reconstructs a denoised signal as a probability-weighted combination. Supervision is provided by a proxy-task model under a collaborative optimization scheme that relies solely on task labels. The compatibility of the task-oriented learning framework with multiple decomposition strategies and network backbones indicates broad applicability. These properties enable practical denoising without manual intervention or clean references and suggest an impact for EEG-based brain–computer interfaces and neuroscience research. Moreover, the task-oriented learning framework may be extended to other bio-signal modalities for denoising. Future work includes validation in online settings and across additional EEG paradigms.

{\small
\bibliographystyle{IEEEtran}
\bibliography{mybib}
}

\end{document}